\documentclass[twocolumn]{aastex701}

\usepackage{
graphicx, 
xcolor, 
bm,
amsmath, 
amssymb,
enumitem
}

\newcommand{\gammasyn}{\gamma_{\rm syn}^\pm}

\newcommand{\gammahillas}{\gamma_H^i}
\newcommand{\betarec}{\beta_{\rm rec}}

\definecolor{haykcolor}{HTML}{bf2844}

\newcommand{\fixed}[1]{{#1}}

\graphicspath{{./figures/}}

\begin{document}

\title{Reconnection-driven Flares in M87*: Proton-Synchrotron-powered GeV Emission}

\correspondingauthor{Hayk Hakobyan}

\author[orcid=0000-0001-8939-6862]{Hayk Hakobyan} 
\affiliation{Physics Department \& Columbia Astrophysics Laboratory, Columbia University, New York, NY 10027, USA}
\affiliation{Computational Sciences Department, Princeton Plasma Physics Laboratory (PPPL), Princeton, NJ 08540, USA}
\affiliation{Center for Computational Astrophysics, Flatiron Institute, 162 Fifth Ave., New York, NY 10010, USA}
\email[show]{haykh.astro@gmail.com}

\author[orcid=0000-0001-7572-4060]{Amir Levinson}
\affiliation{The Raymond and Beverly Sackler School of Physics and Astronomy, Tel Aviv University, Tel Aviv 69978, Israel}
\email{levinson@tauex.tau.ac.il}

\author[orcid=0000-0002-1227-2754]{Lorenzo Sironi}
\affiliation{Physics Department \& Columbia Astrophysics Laboratory, Columbia University, New York, NY 10027, USA}
\affiliation{Center for Computational Astrophysics, Flatiron Institute, 162 Fifth Ave., New York, NY 10010, USA}
\email{lsironi@astro.columbia.edu}

\author[orcid=0000-0001-7801-0362]{Alexander Philippov}
\affiliation{Department of Physics, University of Maryland, College Park, MD 20742, USA}
\affiliation{Institute for Research in Electronics and Applied Physics, University of Maryland, College Park, MD 20742, USA}
\email{sashaph@umd.edu}

\author[orcid=0000-0002-7301-3908]{Bart Ripperda}
\affiliation{Canadian Institute for Theoretical Astrophysics, University of Toronto, Toronto, ON M5S 3H8, Canada}
\affiliation{David A. Dunlap Department of Astronomy, University of Toronto, 50 St. George Street, Toronto, ON M5S 3H4, Canada}
\affiliation{Department of Physics, University of Toronto, 60 St. George Street, Toronto, ON M5S 1A7, Canada}
\affiliation{Perimeter Institute for Theoretical Physics, Waterloo, ON N2L 2Y5, Canada}
\email{bartripperda@gmail.com}







\begin{abstract}
Magnetic reconnection in current layers that form intermittently in radiatively inefficient accretion flows onto black holes is a promising mechanism for particle acceleration and high-energy emission. It has been recently proposed that such layers, arising during flux eruption events, can power the rapid TeV flares observed from the core of M87. In this scenario, inverse Compton scattering of soft radiation from the accretion flow by energetic electron-positron pairs produced near the reconnection layer was suggested as the primary emission mechanism. However, detailed calculations show that radiation from pairs alone cannot account for the GeV emission detected by the Fermi observatory. In this work, we combine analytic estimates with 3D radiative particle-in-cell simulations of pair-proton plasmas to show that the GeV emission can be naturally explained by synchrotron radiation from protons accelerated in the current sheet. Although the exact proton content of the layer is uncertain, our model remains robust across a broad range of proton-to-pair number density ratios. While protons are subdominant in number compared to pairs, our simulations demonstrate that they can be accelerated more efficiently, leading to a self-regulated steady state in which protons dominate the energy budget. Ultimately, proton synchrotron emission accounts for approximately 5\%--20\% of the total dissipation power. The majority is radiated as MeV photons via pair synchrotron emission, with a smaller fraction emitted as TeV photons through inverse Compton scattering.
\end{abstract}


\section{Introduction} 
\label{sec:intro}
Nearly 3000 active galactic nuclei (AGNs) have been detected by the Fermi Large Area Telescope (LAT) at GeV energies, and over $70$ have been detected at TeV energies.
The vast majority of these $\gamma$-ray-loud AGNs have been classified as blazars, in which the emission is believed to originate
from the relativistic jet pointed toward the observer and is therefore strongly beamed, owing to relativistic Doppler boosting. This
strong beaming  naturally accounts for the rapid large amplitude variability commonly observed in blazars, as well as superluminal motions and other emission characteristics.  
However,  about two dozen $\gamma$-ray-emitting AGNs have been identified as radio galaxies, in which the jets are misaligned with respect to the Earth
and the associated Doppler boosting effects are modest or absent. Remarkably, six of those exhibit TeV emission, with two --- IC-310 and M87 --- also 
showing, at times, strong flares (in TeV) with durations as short as the light-crossing time of the putative black hole horizon \citep{RL18}.  This 
unbeamed rapidly varying TeV emission 
likely originates from the close vicinity of the central black hole, offering a unique probe of violent magnetospheric processes.

At very high energies (VHEs), M87 exhibits a hard featureless photon spectrum consistent with a single power law extending from approximately $300$ GeV up to $10$ TeV \citep{EHT_HE_M87}. At lower energies ($0.1...30$ GeV), Fermi-LAT observations reveal excess emission above the standard power-law model beyond $\sim 10$ GeV, along with variability on month-long timescales \citep{RL18,EHT_HE_M87}. These features suggest the presence of two distinct spectral components in the GeV–TeV range. We argue below that the GeV component is likely produced by synchrotron emission from ions (primarily protons) populating the layer and the magnetosphere during the flare, while the TeV component originates from inverse Compton scattering by accelerated pairs.

There are two potential sites at which the unbeamed variable TeV emission observed in misaligned blazars can be produced: spark gaps
generated intermittently at the base of the magnetically extracted Blandford-Znajek jet \citep{Levinson2000, Neronov2007, Levinson2011, Hirotani2016, Hirotani2016b, Chen2020, Crinquand2020, Crinquand2021, Kisaka2022},
and equatorial current sheets produced
episodically during flux eruption states in magnetically arrested accretion \citep{Hakobyan.etal_2023}. In this paper, we focus on the latter scenario.
In the magnetically arrested disk (MAD) state, poloidal magnetic flux is continuously advected inward by the accretion flow and accumulates near the black hole horizon \citep{Tchekhovskoy.etal_2011}. This buildup periodically triggers rapid flux eruption events. During such events, the accretion flow is expelled at certain azimuthal locations beyond roughly $\sim 10r_g$, revealing an equatorial current layer that sustains the polarity reversal of the jet's magnetic field upstream \citep{Ripperda.etal_2022}. This layer becomes unstable to nonlinear tearing modes and undergoes fast magnetic reconnection throughout the duration of the eruption -- typically lasting several to tens of gravitational light-crossing times. Once the excess magnetic flux has been reconnected, the system returns to a quasi-steady accretion state.

The spectrum emitted by particles accelerated in the reconnection layer depends on the magnetization (the magnetic field enthalpy normalized to the enthalpy of the plasma) and the composition of upstream plasma -- the plasma at the base of the jet~\citep[see, e.g., review by][]{2025arXiv250602101S}. Due to limitations in general relativistic magnetohydrodynamic simulations, the ion and pair densities in the inner accretion flow, particularly at the reconnection site, remain poorly constrained. Nonetheless, it is generally expected that the magnetization upstream of the equatorial current sheet is high. If the magnetization is sufficiently large, particles accelerated in the reconnection layer can reach the synchrotron cooling limit, at which point the cooling rate becomes comparable to that of acceleration. At those energies, most of the dissipated magnetic energy is rapidly converted into synchrotron radiation on timescales much shorter than the system's dynamical time. This, in turn, can trigger prolific pair production, which strongly influences the dynamics of the reconnection process. Indeed, recent analysis by \citet{Hakobyan.etal_2023} suggests that in M87, the upstream magnetization is regulated by \textit{in situ} pair creation near the reconnecting layer \citep[also see][]{2023ApJ...944..173C, 2024JCAP...12..009S}. Despite strong radiative losses, the analysis reveals the development of a hard-power-law distribution of pairs. The resulting synchrotron spectral energy distribution (SED) peaks at photon energies of a few tens of MeV. This value, which corresponds to the emission of particles at the burnoff limit, is insensitive to the strength of the magnetic field and is thus universal for a given species (pairs, in this case). A small fraction of the dissipated energy is also channeled into TeV emission via IC scattering of soft disk photons by the accelerated pairs.

The analysis presented in \cite{Hakobyan.etal_2023} considers a pure pair plasma and neglects the presence of ions. However, in realistic scenarios, the plasma at the reconnection site is expected to contain at least a small fraction of ions (i.e., the unevacuated portion of the accretion disk). This raises the important question of how even a subdominant ion component might influence the reconnection dynamics and the resulting emission signatures. In this work, we show that even a small population of ions can significantly impact the observed spectrum by dominating the GeV photon emission through efficient ion-synchrotron cooling. We also demonstrate that ion energy losses via pion photoproduction are negligible under conditions relevant to the accretion around the M87 black hole. By combining analytic estimates with fully self-consistent 3D radiative particle-in-cell (PIC) simulations of the magnetic reconnection in ion–pair plasmas, we compute the resulting emission spectrum in the MeV–TeV range and show that it naturally accounts for the variable VHE emission observed in M87.

In section \ref{sec:analytics}, we present the details of our analytic model, introducing the important dimensionless scales of the problem, as well as computing the expected luminosity radiated via the proton-synchrotron mechanism during the flare. Some of the assumptions from this section are then directly tested in Section~\ref{sec:simulations}, where we present the results from the radiative 3D PIC simulations, as well as produce the synthetic emission spectra at energies from MeV to tens of TeV. We conclude in Section ~\ref{sec:conclusions}, where we discuss the limitations of the model, as well as prospects for future directions.

\section{Analytic estimates}
\label{sec:analytics}

In this section, we elucidate the characteristic energy scales of the distribution of ions accelerated in a reconnecting current sheet and provide a rough estimate for their synchrotron emissivity. Throughout this paper, we use the words ``ions'' and ``protons'' interchangeably, assuming that most of the baryonic matter both in the disk and the jet is composed of protons. For visual clarity, we use the subscript or superscript ``$i$'' to indicate quantities related to the population of ions, while using ``$\pm$'' for pairs.

The number density of ions is denoted as $n_i$, and is henceforth measured in multiples of the Goldreich-Julian (GJ) density: $n_i =\mathcal{M}_i n_{GJ}  = \mathcal{M}_i \Omega B /2 \pi |e| c$, where $B$ is the characteristic magnetic field strength upstream of the reconnection layer, $\Omega \simeq c/2r_g $ is the angular velocity of a rapidly spinning black hole, and $r_g=GM/c^2$ is its gravitational radius (with $M$ being the mass of the black hole\footnote{For M87*, $M\sim 6.5\cdot 10^9~M_\odot$, with $r_g\sim 65$ AU.}, $G$ being the gravitational constant, $e$ being the electron charge, and $c$ being the speed of light).
%
We assume that the current layer has an effective area $S \approx (\alpha/2) r^2$, where $\alpha\sim 1$ is the characteristic azimuthal angular extent of a powerful flux eruption region, while $r\sim 10 r_g$ is the characteristic radial size of the layer \citep{Ripperda.etal_2022}. The rate of energy dissipation inside such a current sheet due to magnetic reconnection is then given by
\begin{widetext}
\begin{equation}
\label{eq:dissp}
L_{\rm rec} \approx \beta_{\rm rec} c\frac{B^2 S}{4\pi} \sim 7\cdot 10^{43}~\alpha\left(\frac{B}{100~\textrm{G}}\right)^2\left(\frac{r}{10 r_g}\right)^2 ~ \text{erg s$^{-1}$},
\end{equation}
\end{widetext}
where $\beta_{\rm rec}$ is the characteristic rate of reconnection, which for collisionless plasmas -- as will be shown below -- is $\approx 0.1$, and the magnetic field strength is taken to be close to $\sim100$ G\footnote{\fixed{The characteristic strength of the magnetic field we employ here is estimated from the jet power, which is also in rough agreement with the value reported by the \cite{2021ApJ...910L..13E}.}}. We also assume that about half of the advected Poynting flux is dissipated during the reconnection \citep[see, e.g.,][]{2014ApJ...783L..21S, 2014PhRvL.113o5005G, 2015MNRAS.450..183S} and that the flux enters the reconnection region from both sides. In our analytic consideration, we ignore the presence of accelerated pairs; in the real system, due to the strong cooling of pairs, their characteristic Larmor radii are much smaller than those of the energetic ions, meaning the dynamics of the two species is spatially decoupled. A full treatment, based on \emph{ab initio} kinetic simulations, will be given in the next section.

There are three important energy scales to consider regarding the dynamics of ions. The first one is the \textit{Hillas limit} -- the energy at which the acceleration timescale with an electric field of strength $\beta_{\rm rec}B$ is comparable to the global advection timescale, $r/c$. This relation yields
\begin{equation}
\label{eq:hillas-limit-ions}
\gamma_{H}^i \sim 3\cdot 10^{10} \left(\frac{B}{100~\textrm{G}}\right)\left(\frac{r}{10r_g}\right).
\end{equation}

\noindent Here, $m_i$ is the proton mass and $|e|$ is its charge. The second scale is the \textit{mean energy of protons}, $\langle\varepsilon_i\rangle \equiv \langle{\gamma_{i}}\rangle m_i c^2$. In this system, we assume that ions carry most of the dissipated energy, with pairs being subdominant and mostly localized in a thin region around the current layer (we will demonstrate this further in our simulations). Thus, the average energy of ions can be estimated by assuming that the force exerted by their pressure inside the 
current layer, $\nabla P_i$, balances the Maxwell stress $\bm{j}\times\bm{B}$, with $\bm{j}$ being the current density and $\bm{B}$ being the magnetic field (both measured upstream of the layer where the protons dominate the current density). In dimensionless form, this equality reduces to
\begin{equation}
\label{eq:sigmai}
\langle \gamma_{i}\rangle \approx \chi_i\sigma_i \equiv \chi_i\frac{B^2/4\pi}{n_i m_i c^2} \sim 2\cdot 10^{10}\left( \frac{B}{100~\textrm{G}}\right) \chi_i\mathcal{M}_i^{-1},
\end{equation}
with $\sigma_i$ being the ``cold'' ion magnetization, while $\chi_i\approx \mathcal{O}(1)$ is a factor that will be measured from our simulations. As we demonstrate in the next section, the value of this parameter is roughly $\chi_i\sim0.1 ... 0.15$. The third energy scale is the \textit{synchrotron burnoff limit}, obtained by balancing the synchrotron cooling time, 
$t_{\rm syn}(\gamma_i) = 9 m_i^3 c^5 / (4e^4 B^2 \gamma_i)$, with the acceleration time for the ions in the electric field of $E_{\rm rec}\sim \beta_{\rm rec}B$ being $t_{\rm acc}(\gamma_i) \approx \gamma_i m_i c / (|e| E_{\rm rec})$. We then obtain
\begin{equation}
\label{eq:gamma-syn-ions}
\gamma_{\rm syn}^i \approx   7\cdot 10^{9} \left(\frac{B}{100~\textrm{G}}\right)^{-1/2}.
\end{equation}
Since $\gamma_{\rm syn}^i \ll \gamma_{H}^i$, the maximum energy of protons is limited by synchrotron cooling rather than the escape from the system. Additionally, for ion multiplicities $\mathcal{M}_i \gtrsim 3 \left(B/100~\textrm{G}\right)^{3/2}$, the mean energy of protons is smaller than the cooling limit, namely $\langle{\gamma_{i}}\rangle \lesssim \gamma_{\rm syn}^i$. It is important to note here that in the MAD state, magnetization in the disk corresponds to \fixed{$\sigma_i\lesssim 1$} owing to the high density of accretion-disk protons. Assuming the magnetic field strength does not vary too much between the disk and the jet, we see that this implies that the density in the disk is orders of magnitude larger than the GJ value; i.e., compare \fixed{$\sigma_i\lesssim 1$} for $n_{\rm disk}^i$, and \eqref{eq:sigmai} where we took $\mathcal{M}_i\equiv n_i/ n_{\rm GJ}\sim 1$. 
We will thus further assume that a small fraction of this accretion-disk matter --- $\mathcal{M}_i\gg 3$ --- remains in the current layer during the flux eruption event and that the average energy of protons is always much smaller than the radiation burnoff: $\langle\gamma_i\rangle \ll \gamma_{\rm syn}^i$. \fixed{Note also that the amount of pairs in the layer estimated by \cite{Hakobyan.etal_2023} is of the order of $n_\pm\approx 10^6...10^7n_{\rm GJ}$, meaning that our further discussion holds, as long as the number density of protons is much smaller than that value, i.e., $3\ll\mathcal{M}_i\ll10^6...10^7$.}\footnote{\fixed{The upper limit corresponds to $\sigma_i\gg \sigma_\pm m_\pm n_\pm/m_i n_i\gtrsim 500$.}}

\subsection{Synchrotron Emission of Protons}

With this hierarchy of scales --- as will be shown in the next chapter --- the energy distribution of the accelerated protons is a broken power law: 

\begin{equation}
\label{eq:ion-energy-dist}
\frac{dn_i}{d\gamma_i }= n_i^0
\begin{cases}
			 \gamma_i^{-1}, & 1 \le  \gamma_i < \gamma_{b}^i \\
           \left(\gamma_{b}^i\right)^{s-1} \gamma_i^{-s}, & \gamma_{b}^i \le \gamma_i \le \gamma_{\rm syn}^i
		 \end{cases}
\end{equation}

\noindent with a break energy at $\gamma_{b}^i$. We also assume that the distribution exponentially cuts off beyond $\gamma_{\rm syn}^i$. The plasma kinetic simulations described in the next section indicate that the power-law index $s \approx 2$. The constant $ n_i^0$ is related to the number density via
\begin{equation}
\label{eq:ion_density}
n_i \equiv \int \frac{dn_i}{d\gamma_i} d\gamma_i =  n_i^0 \underbrace{\left[\frac{1}{s-1}\left(1-\left\{\frac{\gamma_b^i}{\gamma_{\rm syn}^i}\right\}^{s-1}\right)+\ln \gamma_{b}^i \right]}_{\chi_n},
\end{equation}
where $15 \lesssim \chi_n \lesssim 20 $ for values of $\gamma_b^i$ in the range $10^6 \le \gamma_b^i \le 10^9 $ (assuming $s>1$, and $\gamma_b^i\ll \gamma_{\rm syn}^i$). The total energy density of the accelerated protons is 
\begin{equation}
\begin{split}
\frac{U_i}{m_i c^2} & =\int \gamma_i\frac{dn_i}{d\gamma_i} d\gamma_i \\
&=  n_i^0 \gamma_{b}^i \underbrace{\begin{cases}
1-1/\gamma_b^{i}+\frac{1}{s-2}\left(1 - \left\{\frac{\gamma_b^i}{\gamma_{\rm syn}^i}\right\}^{s-2}\right),\quad &s\ne 2,\\
1-1/\gamma_b^{i}+\ln{\frac{\gamma_{\rm syn}^i}{\gamma_b^i}},\quad &s=2.
\end{cases}}_{\chi_U}
\end{split}
\end{equation}

\noindent The break energy is then related to the mean energy of protons, $\langle{\gamma}_i\rangle = U_i/n_i m_i c^2 = \chi_i \sigma_i$, via $\gamma_{b}^i = \sigma_i (\chi_i\chi_n/\chi_U)$; formally, this is an implicit equation on $\gamma_{b}^i$.

Adopting $\nu_L\equiv e B/2\pi m_i c$, we can estimate the ion-synchrotron emissivity, $4\pi j_\nu$, in the frequency interval $\left(\gamma_{b}^i\right)^2 < \nu/(3\nu_L/2) < \left(\gamma_{\rm syn}^i\right)^2$ \citep{Blumenthal.Gould_1970}:



\begin{equation}
\label{eq:emissivity}
4\pi j_\nu = 4\pi a(s)\frac{n_0^i |e|^3 B}{m_i c^2}\left(\gamma_b^i\right)^{s-1}\left(\frac{\nu}{\frac{3}{2}\nu_L}\right)^{(1-s)/2},
\end{equation}

\noindent where $a(s)$ is a combination of $\Gamma$ functions with $a(2)\approx 0.1$. Integrating the above equation from $\nu_b\equiv (3/2)\nu_L\left(\gamma_{b}^i\right)^2$ to $\nu_{\rm syn}\equiv (3/2)\nu_L\left(\gamma_{\rm syn}^i\right)^2$ over the volume $V$ yields the total emitted power. For the volume, we may take $V=S \delta_i$, where $S$ is the effective surface area of the layer, and $\delta_i$ is the characteristic thickness of the region where most of the radiation is emitted. For the latter, we take the Larmor radii of particles close to the synchrotron burnoff limit, i.e., $\delta_i =\gamma_{\rm syn}^i m_i c^2/|e|B$. Substituting $n_i^0\equiv n_i/\chi_n$ from Equation \eqref{eq:ion_density}, $(\gamma_{\rm syn}^i)^2\equiv (9/4)\beta_{\rm rec}m_i^2 c^4/|e|^3 B$, $\gamma_b^i=\sigma_i\chi_i\chi_n/\chi_U$, and $\sigma_i=B^2/4\pi n_i m_ic^2$, and performing the integration over frequencies (assuming $\gamma_{\rm syn}^i\gg \gamma_b^i$, and $s<3$), we arrive at the total synchrotron power emitted by the ions:

\begin{equation}
\label{eq:L_syn}
\begin{split}
L_{\rm syn}^i &\approx (3/2)\nu_L V\int_{\left(\gamma_b^i\right)^2}^{\left(\gamma_{\rm syn}^i\right)^2}4\pi j_\nu d\left\{\frac{\nu}{(3/2)\nu_L}\right\} \\
 & \approx \frac{27a(s)}{6-2s}\frac{\chi_i}{\chi_U}\underbrace{\beta_{\rm rec}c\frac{B^2 S}{4\pi}}_{L_{\rm rec}}\left(\frac{\gamma_{\rm syn}^i}{\gamma_b^i}\right)^{2-s}.
\end{split}
\end{equation}

\noindent The ratio $L_{\rm syn}^i/L_{\rm rec}$ is thus the fraction of the dissipated energy emitted as ion-synchrotron. For $s=2$, $B\approx 100$ G, the value of $\chi_U$ varies only between $3...10$ with $\gamma_b^i \approx 10^{6...9}$. For characteristic values of $\chi_U\sim 5$, and $\chi_i\sim 0.1...0.2$, we thus estimate that about $5...10\%$ of the total dissipated power during reconnection can be emitted by the protons as synchrotron emission. This emission will peak at around $\nu_{\rm syn}\equiv(3/2)\nu_L \left(\gamma_{\rm syn}^i\right)^2$, which, by the definition of $\gamma_{\rm syn}^i$, does not depend on the magnetic field strength and roughly corresponds to energies of $\sim 40$ GeV.  

For $s>2$ the radiative efficiency is considerably smaller. In practice, it means that in this case, in the absence of the guide field, ions will continue to accelerate until $s \approx 2$ is reached. This is indeed what is seen in the simulations. Note that our argument implicitly assumes that the ions accelerate in the weak-cooling regime, such that $\gamma_b^i\lesssim \sigma_i\ll \gamma_{\rm syn}^i$, which in turn requires high-enough multiplicity. In cases where the ion density is small ($n_i\lesssim 3 n_{ GJ}$), and $\{\sigma^i,\gamma_b^i\}\gtrsim\gamma_{\rm syn}^i$, the acceleration will proceed until ions reach the burnoff limit, $\gamma_{\rm syn}^i$, maintaining a relatively hard --- $f_i\propto\gamma_i^{-1}$ --- power-law slope. The corresponding synchrotron spectrum will scale as $4\pi j_\nu\propto \nu$ all the way to $\sim 40$ GeV (which corresponds to $\gamma_i\sim\gamma_{\rm syn}^i$). Such a spectrum is inconsistent with the observations, disfavoring extremely low ion multiplicities.

Importantly, in our analytic considerations, we neglected the dynamics of pairs, which are assumed to be mainly confined to a thin region in the midplane, due to strong cooling (as will be shown below). This assumption is well justified for realistic parameters, as the ratio between the Larmor radii of pairs and protons --- both computed at their respective synchrotron burnoff Lorentz factors --- is $r_L^i(\gamma_{\rm syn}^i)/r_L^\pm(\gamma_{\rm syn}^\pm)\approx (m_i/m_\pm)^2\gg 1$.

\subsection{Drag Due to Pion Photoproduction}
The accelerated protons will also lose energy through pion photoproduction upon collision with soft photons emitted by the accretion flow.
The threshold photon energy, $\varepsilon_{\rm \Delta}$, for which head-on collision with a proton of maximum energy $\varepsilon^i_{\rm max}= m_i c^2 \gamma^i_{\rm syn}$ (eq.~\ref{eq:gamma-syn-ions})
is at the $\Delta$-resonance is $\varepsilon_{\rm \Delta} \approx 10^{-2}$ eV, roughly the observed SED peak energy.
Since during large-flux-eruption states the disk recedes to beyond $\sim 10 r_g$, we anticipate that the soft radiation will emerge from radii $r_s \gtrsim 10 r_g$.  In what follows, we adopt soft photon luminosity of $L_{s} \sim 10^{42}$ erg s$^{-1}$
and mean soft photon energy $\varepsilon_s \gtrsim \varepsilon_\Delta$. The corresponding photon density is 

\begin{equation}
\begin{split}
n_{s} &\approx \frac{L_s}{4\pi cr^2_{s} \varepsilon_s} \sim 10^{12} 
\,\textrm{cm}^{-3} \\
&\times
\left(\frac{L_s}{10^{42}~\textrm{erg}~\textrm{s}^{-1}}\right) 
\left(\frac{r_s}{10 r_g}\right)^{-2}
\left(\frac{\varepsilon_s}{0.01\, \textrm{eV}}\right)^{-1} 
.
\end{split}
\end{equation}

\noindent The energy loss rate due to pion photoproduction is given by $t^{-1}_{p\gamma} \approx \kappa_{p\gamma}\sigma_{p\gamma}n_{s} c$,
where $\kappa_{p\gamma}$ is the inelasticity factor, and $\sigma_{p\gamma}$ is the characteristic cross section. Adopting $\kappa_{p\gamma}\sigma_{p\gamma} \approx 0.05$ mb, one finds
\begin{equation}
\label{eq:t_pg-disk}
\begin{split}
t_{p\gamma}^{-1} &\sim 2\cdot 10^{-6} ~\text{s}^{-1}\\
&\times\left(\frac{L_s}{10^{42}~\textrm{erg}~\textrm{s}^{-1}}\right) 
\left(\frac{r_s}{10 r_g}\right)^{-2}
\left(\frac{\varepsilon_s}{0.01\, \textrm{eV}}\right)^{-1}
.
\end{split}
\end{equation}
Equating the timescale from Equation \eqref{eq:t_pg-disk} with the acceleration rate in reconnection, $t^{-1}_{\rm acc}(\gamma_i) \approx |e|\beta_{\rm rec} B/ \gamma_i m_i c $,
yields a maximum Lorentz factor:
\begin{equation}
\begin{split}
\gamma^i_{p\gamma} &\approx 5\cdot 10^{10} \\ 
\times
&\left(\frac{B}{100~\textrm{G}}\right)
\left(\frac{L_s}{10^{42}~\textrm{erg}~\textrm{s}^{-1}}\right)^{-1}
\left(\frac{\varepsilon_s}{0.01\, \textrm{eV}}\right)
\left(\frac{r_s}{10 r_g}\right)^{2},
\end{split}
\end{equation}
which exceeds the synchrotron burnoff limit given by Equation \eqref{eq:gamma-syn-ions} by an order of magnitude. We thus conclude that for energies below the synchrotron burnoff limit, energy losses of protons due to pion photoproduction are negligible compared to synchrotron losses. 

\section{PIC Simulations}
\label{sec:simulations}

\begin{figure*}[htb!]
\centering
\includegraphics[trim=800 300 0 200,clip,width=0.8\textwidth]{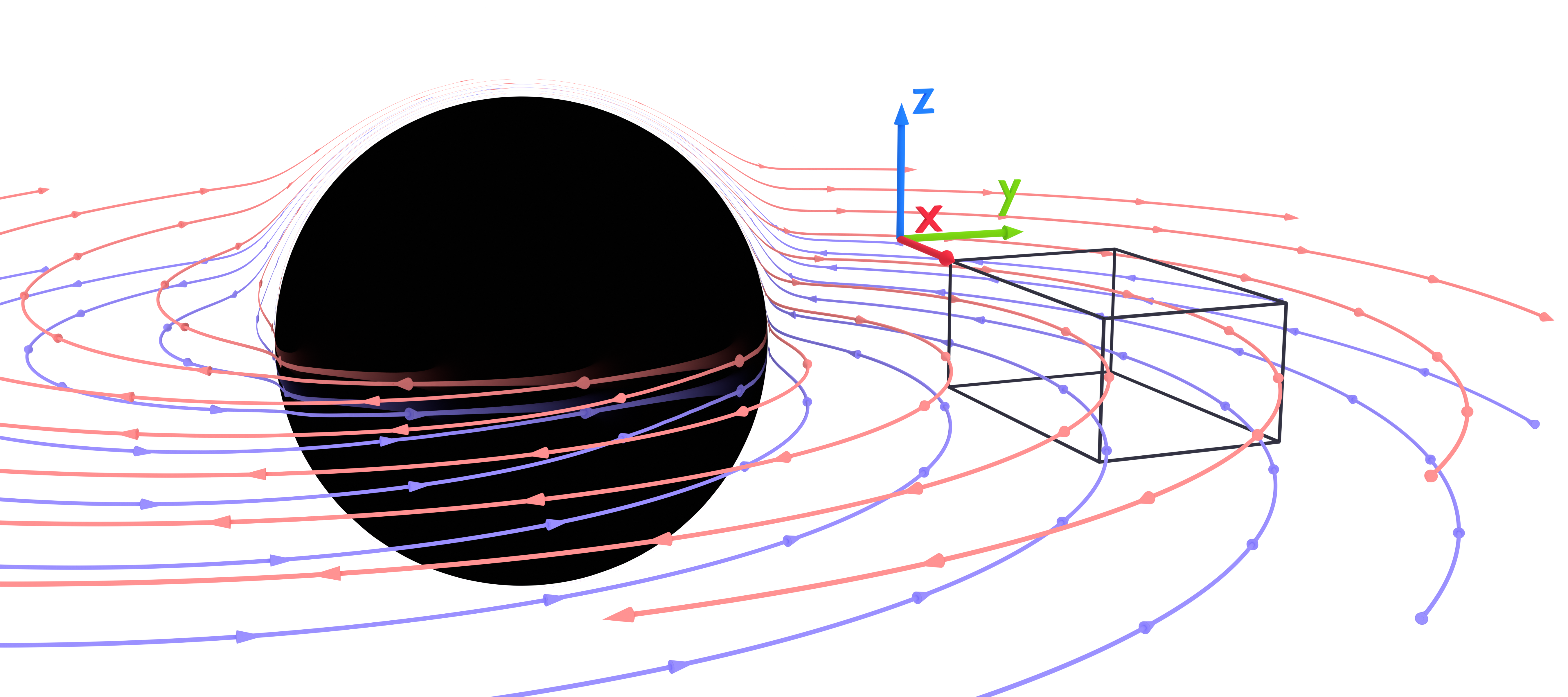}
\caption{Schematic illustration of the simulation box (the black rectangle) near the exposed equatorial current layer around the black hole. Toroidal magnetic field lines of different polarities are sketched with the red and blue arrows. The colored $x$-, $y$-, and $z$-axes correspond to the actual orientations of the axes in our simulation. Note that the geometry of the layer implies the absence of a guide magnetic field (in this case, a component along the $y$-axis).}
\label{fig:3dscheme}
\end{figure*}

In this section, we study the dynamics of electron-positron-ion plasma during relativistic magnetic reconnection (see Figure~\ref{fig:3dscheme}).\footnote{Note, that our setup focuses on an isolated current layer without considering the global dynamics of the accretion flow.} Our main goals for the following subsections will be to:

\begin{enumerate}[label=(\alph*)]
    \item establish the efficiency of ion acceleration and their resulting energy distribution during pair-dominated reconnection and evaluate the dimensionless parameter, $\chi_i$, that we used in our analytic model in Section~\ref{sec:analytics};
    \item understand the feedback of the accelerated ions on the reconnecting layer, which will allow us to extrapolate our results to realistic parameters; and
    \item reconstruct a realistic spectrum of high-energy photons during the flare using the pair+ion-synchrotron emission, as well as the IC emission of pairs.
\end{enumerate}

\subsection {Setup \& Dimensionless Parameters}
\label{subsec:setup_params}

We use 3D radiative PIC simulations of a localized current layer using the multispecies \texttt{Tristan v2} PIC code \citep{2023zndo...7566725H}. The entire domain has an extent of $L\times L\times 0.8 L$, where the magnetic field upstream points in $\hat{\bm{x}}$, with the layer occupying the $x\textrm{-}y$ plane (see Figure~\ref{fig:3dscheme}). In our simulations, we resolve $L$ with $2000$ grid cells. We start with a Harris equilibrium with no guide field, with the magnetic field upstream being $\bm{B}=B_\circ \hat{\bm{x}}\tanh{\left(z/\Delta_{\rm cs}\right)}$; here, $B_\circ$ corresponds to the magnetic field upstream, while $\Delta_{\rm cs}$ is the thickness of the initial layer. The domain is initially filled with an electron-positron-ion plasma of total number density $n_i^\circ + n_\pm^\circ = n_\circ$. Both the mass and the number of particles in our simulations are always dominated by pairs, with the ions comprising a small fraction of all particles: $f_i \equiv n_i^\circ/(n_i^\circ + n_+^\circ)\ll 1$ and $m_in_i\ll m_\pm (n_++n_-)$ (note that $n_i+n_+=n_-$). To make the separation of scales tractable, we also reduce the mass ratio \fixed{$\mu_i \equiv m_i/m_\pm = 5$} in all of our simulations, although our conclusions are insensitive to the exact value of this parameter \citep[see][]{Chernoglazov.etal_2023}.\footnote{This is true as long as there is strong synchrotron cooling of one of the positively charged species, and not the other, since in relativistic dynamics, the inertia of particles is dictated by their ``effective'' mass, $\gamma m$. \fixed{We have confirmed these results by running simulations with smaller, $\mu_i=1$, and bigger, $\mu_i=10$, values.}} Boundaries in the $x\textrm{-}z$ plane are periodic, while in the $x\textrm{-}y$ plane we replenish the upstream plasma and the fields, absorbing any outgoing electromagnetic waves and particles, and in the $y\textrm{-}z$ plane we impose outflow boundaries on both the fields and the particles. To facilitate the largest possible separation of scales (as discussed below), and to be able to run the simulation for long enough, we set the characteristic number of particles per cell upstream (corresponding to $n_\circ$) to $2$, and resolve the fiducial skin depth: $d_\circ\equiv \sqrt{m_\pm c^2/(4\pi n_\circ e^2)}$ with two simulation cells. As discussed by \citealt{Chernoglazov.etal_2023}, the resulting dynamics of high-energy particles is well preserved with this choice of sample size and resolution.

Since we are interested in the ultrarelativistic limit, we set $\sigma_\circ = B_\circ^2/(4\pi n_\circ m_\pm c^2) = 100$ for all of our simulations.\footnote{Notice that this value does not exactly correspond to the actual upstream magnetization, because of the presence of ions. However, because of their low number density, $\sigma_\circ \approx B_\circ^2/(4\pi (n_\pm^\circ m_\pm + n_i^\circ m_i)c^2)$.} The equation of motion for pairs also contains a synchrotron drag force, using the algorithm introduced by \cite{Tamburini.etal_2010}. Additionally, since we are interested in the dynamics of ions, we also define the \emph{ion magnetization}, $\sigma_i\equiv B^2/(4\pi n_im_ic^2)$, which can be expressed as $\sigma_i=\sigma_\circ(2/f_i-1)/\mu_i$. The strength of synchrotron losses for the pairs is parameterized with the critical Lorentz factor, $\gamma_{\rm syn}^\pm$, \citep[see, e.g.,][]{2014ApJ...780....3U}, which is defined through the following relation: $0.1 B_\circ |e| \equiv (\sigma_T/4\pi) B_\circ^2 \left(\gammasyn\right)^2$. In all of our simulations, we pick $\gammasyn<\sigma_\circ$, which corresponds to the strong-cooling regime; namely, pairs start cooling faster than they accelerate (in the electric field of strength $0.1 B_\circ$) before they reach the energy of $\sigma_\circ m_\pm c^2$ (this is only true for pairs moving perpendicular to upstream magnetic field). Note that the value of magnetization we employ in our simulations, as well as the synchrotron burnoff limit for pairs, are both much smaller than the actual values near the black hole accretion flow. Despite this, the essential dynamics of the system is accurately captured, as long as the hierarchy of dimensionless scales --- in this case $\sigma_\circ$ and $\gamma_{\rm syn}^\pm$ --- is preserved.

Pairs dominate the number density; however, their energy density is limited by synchrotron cooling to around $n_\pm \gammasyn m_\pm c^2$ \citep{Hakobyan.etal_2023}. Ions, on the other hand, are uncooled in our simulations and are thus free to accelerate to arbitrarily high energies.\footnote{In the real system, protons do experience synchrotron cooling, but it is negligible compared to their acceleration until they reach the burnoff limit, $\gamma_{\rm syn}^i$. Our results are thus applicable only to energies below this burnoff limit, which is also the range where most of the proton-synchrotron emission will be generated.} At some point, as the energy content of ions grows, they become dynamically important, as their energy density (or, equivalently, pressure) becomes comparable to that of the upstream magnetic field. Our choice of parameters --- in particular $f_i\ll1$, and $\gamma_{\rm syn}^\pm\leq \sigma_\circ$ --- ensures that the characteristic ion Larmor radii, $\sigma_i m_i c^2/|e|B_\circ \approx 2\sigma_\circ/f_i (m_\pm c^2/|e|B_\circ)$, are significantly larger than the scale height of the pair-dominated current layer, $\gamma_{\rm syn}^\pm (m_\pm c^2/|e|B_\circ)$, providing sufficient decoupling between the high-energy protons and the pairs.
Additionally, ions will be unable to accelerate if their Larmor radii reach the size of the box: $\gammahillas m_i c^2 / |e| B_\circ \approx L$, where we will refer to the critical Lorentz factor $\gammahillas$, as the Hillas limit (we employ $\gammahillas\gtrsim \sigma_i$). Ultimately, the typical hierarchy of dimensionless scales we maintain in this work is the following:

\begin{equation}
\begin{aligned}
    &\textrm{strong pair-cooling: }\gammasyn\ll\sigma_\circ, \\
    &\textrm{decoupling of ion scales: }\gammasyn\ll \underbrace{2\sigma_\circ/f_i}_{\approx\mu_i\sigma_i}\lesssim\gamma_H^i.
\end{aligned}
\end{equation}


\subsection {Energy Partition}

\begin{figure}[htb!]
\includegraphics[width=\columnwidth]{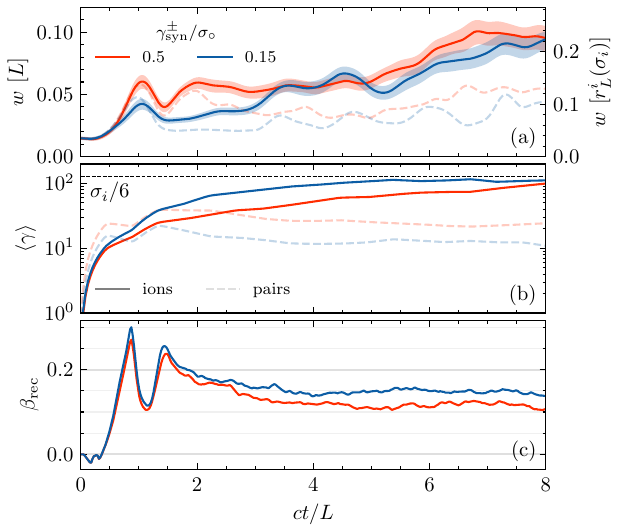}
\caption{Time evolutions of space-averaged quantities for two simulations with marginal (red; $\gammasyn/\sigma_\circ= 0.5$) and strong (blue; $\gammasyn/\sigma_\circ=0.15$) synchrotron cooling of pairs. (\emph{a}): Width of the layer along $z$ measured as the FWHM of the energy density (averaged in $x$ and $y$) of pairs (dashed) and protons (solid); the shaded region corresponds to about $25\%$ variation in the FWHM value, indicating how steep the gradient of $w(z)$ is. (\emph{b}): Mean Lorentz factor of pairs (dashed) and protons (solid), that have participated in reconnection. (\emph{c}): Space-averaged dimensionless reconnection rate measured as $(\bm{E}\times\bm{B})_z/B^2$ in the region $0.1<z<0.2$. Note that in the case with stronger cooling (blue), the pressure inside the plasmoids is partially provided by the out-of-plane magnetic field. Thus, the width of the pair-dominated region in (\emph{a}) at later times is similar for both strong and weak cooling.}
\label{fig:meangamma_betarec_vs_t}
\end{figure}

In this subsection we present the results from two simulations with $\mu_i = 5$, $f_i = 0.05$, $\sigma_\circ = 100$, and $\gammasyn = \{15,50\}$, with the size of the box being $L=2016$ cells; the ion magnetization value is thus $\sigma_i= 780$ in both simulations, while the Hillas limit for the protons is $\gamma_H^i\approx 2000$. We pick two values for the pair-synchrotron burnoff to ensure our conclusions about the acceleration of protons do not depend on the dynamics of pairs --- the assumption we made in Section~\ref{sec:analytics}. In Figure~\ref{fig:meangamma_betarec_vs_t} we present the time evolutions of several volume-averaged quantities from both of our simulations. Quantities from the weaker-cooling run, $\gammasyn/\sigma_\circ=0.5$ ($\gammasyn=50$), are shown in red, while those for the stronger cooling are shown in blue. In panel (\emph{a}), we show the characteristic width of the layer, $w$, evaluated as the FWHM along the $z$-direction of the integrated (in both $x$ and $y$) proton energy density (solid curves) and the pair energy density (dashed curves). Panel (\emph{b}) shows the mean Lorentz factor of both protons (solid) and pairs (dashed) accelerated in the layer. Since the distribution of protons at late times is a broken power law with $f_i\propto \gamma^{-1}$ below a break (as demonstrated later), their mean energy is a good measure for the break itself. The dashed horizontal line indicates the energy of the break, $\gamma_b^i$, as a fraction of $\sigma_i$ (the exact value of this break is discussed further in Section~\ref{sec:proton_feedback}).

The average reconnection rate, measured as $\beta_{\rm rec}=(\bm{E}\times\bm{B})_z/B^2$, is shown in panel (\emph{c}). After a brief transient lasting about $2L/c$, a steady state is established, during which the dissipated magnetic energy is deposited into the kinetic energy of pairs and protons. The mean energy of pairs is quickly saturated at a value comparable to the corresponding $\gammasyn$ (the dashed lines in panel (\emph{b})), as they constantly radiate their energy via synchrotron emission. Ions, on the other hand, keep gaining energy, with their mean Lorentz factor growing with time to about a fraction of $\sigma_i$. Likewise, the characteristic width of the layer determined by the pressure (energy density) of ions increases, proportional to their characteristic Larmor radii, $\langle\gamma_i\rangle m_ic^2/|e|B_\circ$ (panel (\emph{a})). Note, that beyond $ct/L\gtrsim 4$, the width of the layer occupied by the accelerated ions, regardless of the degree of pair cooling, is fully decoupled from the pair-dominated current layer, the width of which is shown with the dashed lines in panel (\emph{a}). 



As we demonstrate further, the main acceleration channel for protons is similar to the one discovered by \cite{Zhang2021, Zhang2023} for pairs and studied extensively by \cite{Chernoglazov.etal_2023}. In this scenario, ions are demagnetized from the main layer, after being initially energized at the X-point, and enter the ``free'' acceleration stage upstream, where they tap the global ideal electric field $E_y\approx \betarec B_\circ$. Note that the saturated mean energy of ions is almost the same in both cases $\langle\gamma_i\rangle\sim \sigma_i/6$ (regardless of the strength of the pair cooling), and is $\langle\gamma_i\rangle \ll \gamma_H^i$. This separation of scales implies that protons have the capacity to accelerate further; however, when the threshold is reached, the acceleration with $f_i\propto \gamma^{-1}$ slows down, and the distribution steepens to $f_i\propto \gamma^{-2}$ establishing a steady state (as shown in Section~\ref{subsec:energydist}). This indicates that ions become dynamically important and feed back into the system, effectively inhibiting further increase in their mean energy; we study this feedback effect of the following section. 


Before moving on further, let us also note that the reconnection rate is insensitive to the proton feedback, as is evident by the fact that $\beta_{\rm rec}$ is constant at late times of our simulation, when the mean ion energy no longer evolves. As shown in Figure~\ref{fig:meangamma_betarec_vs_t}(\emph{c}) at $ct/L\gtrsim 4$, the rate establishes a value of around $\betarec\approx 0.1...0.15$ in the steady state, with no significant variations. This should come as no surprise, as inhibiting the reconnection rate would have caused ions to accelerate slower, thus weakening their feedback. Notice also that both the mean energy of the ions (panel (\emph{b})), as well as the reconnection rate (panel (\emph{c})) are slightly higher for the case where pairs are cooled faster $\gammasyn/\sigma_\circ=0.15$ (blue curves). This result is in agreement with \cite{Chernoglazov.etal_2023} and is likely due to the pair-dominated plasmoids occupying smaller surface area of the sheet, as synchrotron cooling removes their pressure support, effectively compressing them. This, in turn, means that: (1) protons are less likely to get captured by the plasmoids, which would inhibit their ``free'' acceleration; and (2) more X-points can be formed, leading to a higher influx rate and thus a stronger electric field.

\subsection{Proton Feedback}
\label{sec:proton_feedback}

\begin{figure*}[htb!]
\includegraphics[width=\textwidth]{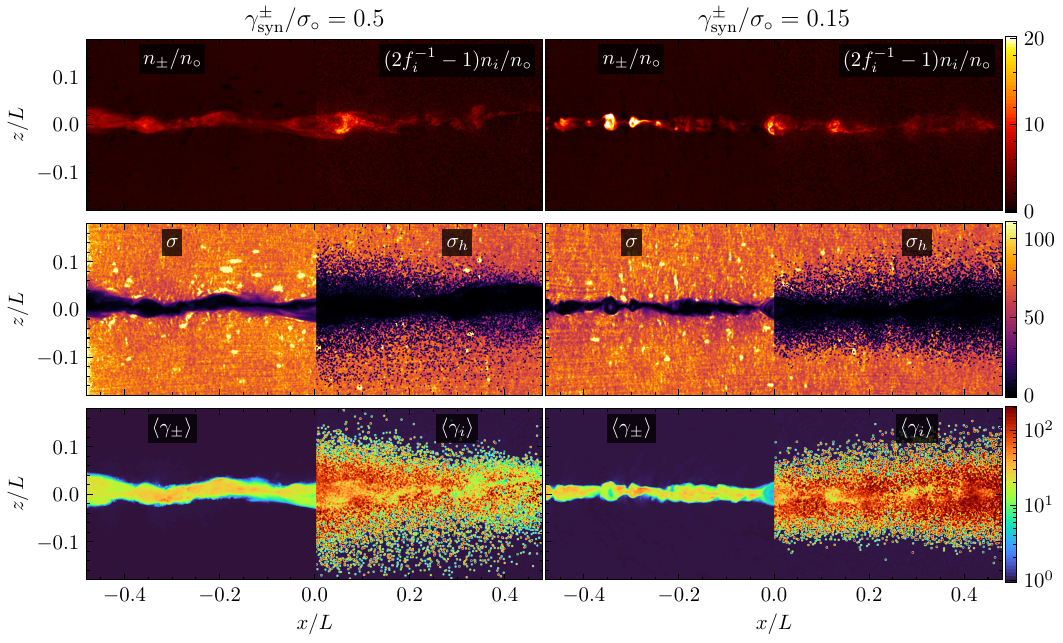}
\caption{Snapshots from two different simulations, with the columns corresponding to different values for the synchrotron cooling strengths of pairs, as indicated in the headings. Each panel is split into two: on the left, we show all the quantities related to pairs, while on the right, we show the quantities related to protons. The first row shows the number densities of pairs and protons (the number density of protons is compensated by their upstream ratio). In the second row we show the cold (left half) and the hot (right half) magnetization parameters, where the role of ions is clearly emphasized. The final row shows the mean Lorentz factors. These plots demonstrate a clear separation between the cooled pairs confined within the plasmoids and the hot uncooled protons that supply most of the pressure farther upstream.}
\label{fig:slice-snapshots}
\end{figure*}

Once the region occupied by the uncooled protons --- which constitute only a small fraction of all the particles in our simulation --- reaches a certain characteristic width, their further acceleration with the $f_i\propto \gamma^{-1}$ slope is inhibited, as they become dynamically important and start backreacting on the reconnection process. From Figure~\ref{fig:meangamma_betarec_vs_t}a, this critical width is about $\approx 0.15...0.2r_L^i(\sigma_i)$, where $r_L^i(\sigma_i)\equiv \sigma_i m_ic^2/|e|B_\circ$. To demonstrate more clearly what the reconnection sheet looks like during this stage, in Figure~\ref{fig:slice-snapshots} we show 2D slices from both of our simulations at a late stage ($ct/L\approx 7$). The two sides of each panel ($x<0$ and $x>0$) depict two different quantities. The top row shows the number densities of pairs and protons (compensated by their initial number density ratio). The middle row shows the cold, $\sigma\equiv B^2/(4\pi(\rho_\pm+\rho_i)c^2)$, and the hot, $\sigma_h\approx B^2/(4\pi(U_\pm+U_i))$, magnetizations (with $\rho$ and $U\approx \langle\gamma \rho\rangle c^2$ being the mass density and the energy density, respectively). In the bottom row, we show the mean Lorentz factor of pairs and ions. 

The density plots of the top row show that the protons on average follow almost exactly the pairs, with their number density ratio being roughly constant throughout the box. This should come as no surprise, as charge neutrality has to be satisfied. Nonetheless, from the middle and bottom rows, it is evident that the dynamics of the ions is very different from that of the pairs. In particular, protons carrying most of the energy density (with $\gamma_i\gtrsim 0.1\sigma_i$) have Larmor radii exceeding the typical sizes of pair-dominated plasmoids and are thus free to escape upstream. Moreover, the amount of energy in protons is enough to reduce the effective hot magnetization significantly (with respect to its value far upstream), meaning that upstream $U_i$ reaches a fraction of $B^2/8\pi$. The main difference between the two cooling cases (left and right columns) is the geometry of the pair-dominated layer, which in the strong-cooling case (right column) is thinner, thus providing a larger separation of scales between the energetic ions and the cooled pairs. Note that the pressure within the plasmoids is still marginally dominated by pairs (by a factor of $\sim 2...3$), as will be demonstrated further. However, the contribution of pairs quickly drops to zero outside the plasmoids, as their Larmor radii are small, due to the strong-cooling losses.

\begin{figure}[htb!]
\includegraphics[width=\columnwidth]{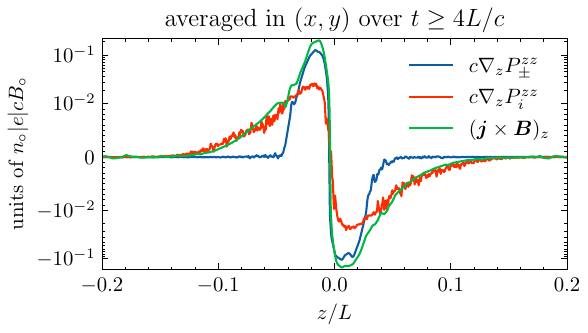}
\caption{Time- and $x$-$y$-averaged forces plotted against the $z$-coordinate (perpendicular to the layer) for the $\gammasyn/\sigma_\circ=0.15$ simulation. The green line shows the magnetic tension force acting toward $z=0$ attempting to compress the sheet. The blue and red lines show the pressure gradients of pairs and protons, respectively, acting in opposition to the $\bm{j}\times\bm{B}$ force. Outside $|z|\gtrsim 0.02$, the magnetic tension is balanced primarily by the pressure of ions.}
\label{fig:pressure}
\end{figure}

The separation of the most energetic protons from the pair-dominated layer introduces an extra pressure component in $z$, and an additional scale much larger than the thickness of the much smaller pair layer. This additional much-more-spread-out pressure in the region of $0.05\lesssim|z|/L\lesssim 0.1$ must be balanced by electromagnetic stresses in $y$, as $c\nabla_i P^{iz} = \left(\bm{j}\times\bm{B}\right)_z$. Here, $P^{ij} \equiv \sum_s m_s c^2 \int f_s \left(u^i u^j / u^0\right) d^3\bm{u}$, where the integral is taken in the comoving (Eckart) frame, $f_s$ is the distribution function for species $s$, while $u^i$ and $u^0$ are the spatial and temporal components of the dimensionless four-velocity. To demonstrate this, in Figure~\ref{fig:pressure} we plot contributions to the pressure gradient from both protons and pairs and compare with the $(\bm{j}\times\bm{B})_z$ term (orange line) for our strongly cooled simulation ($\gammasyn/\sigma_\circ=0.15$). The values are plotted against the $z$-coordinate (with $z=0$ corresponding to the midplane) and are average both in space, $(x,y)$, and time. In the core of the pair-dominated layer, $|z|\lesssim 0.02L$, the pressure gradient is dominated by pairs (blue line). However, at larger distances from the sheet, pairs can no longer contribute, since they are effectively trapped, and the contribution of protons to the pressure gradient takes over (red line). 

To see how $\sigma_i$ enters the balance equation, we may look at Amp\`{e}re's law in the upstream region in a steady state: $(4\pi/c)\bm{j}\approx \nabla\times\bm{B}$. Consider the $y$-component of this equation in the region $0.02L\lesssim |z|\lesssim 0.1L$: $(4\pi/c)j_y \approx (\nabla\times \bm{B})_y\approx \partial_z B_x$. The current density in that region, $j_y$, is provided primarily by the unmagnetized ions $j_y \approx \nu_i \langle\beta_y^i\rangle n_i|e|c$, where $\nu_i\sim\mathcal{O}(1)$ is the fraction of free-streaming ions and $\langle\beta_y^i\rangle\approx 1$ is their average three-velocity perpendicular to the reconnecting magnetic field. In a steady state, the scale height of $B_x$ is proportional to $\langle r_L^i\rangle\equiv \langle\gamma_i\rangle m_i c^2/|e|B_\circ$, and we can parameterize this by assuming $(\nabla\times\bm{B})_y\approx B_\circ/\xi_L r_L^i$, where $\xi_L$ is another unknown dimensionless parameter. Rewriting this relation, and using the definition for $\sigma_i \equiv B_\circ^2 / 4\pi n_im_i c^2$, we thus find

\begin{equation}
    \langle\gamma_i\rangle \approx \sigma_i \underbrace{\frac{\xi_L \langle\beta_y^i\rangle}{\nu_i}}_{\chi_i}.
\end{equation}

The combination of unknown dimensionless values we employ now transparently shows how the unknown coefficient $\chi_i$ is constructed. Its exact value, defined in Section~\ref{sec:analytics}, depends on the distribution and time-averaged trajectories of free ions, which supply both the current and the pressure in the upstream region, establishing equilibrium.



\subsection{Energy Distribution of Pairs \& Ions}
\label{subsec:energydist}

\begin{figure}[]
\includegraphics[width=\columnwidth]{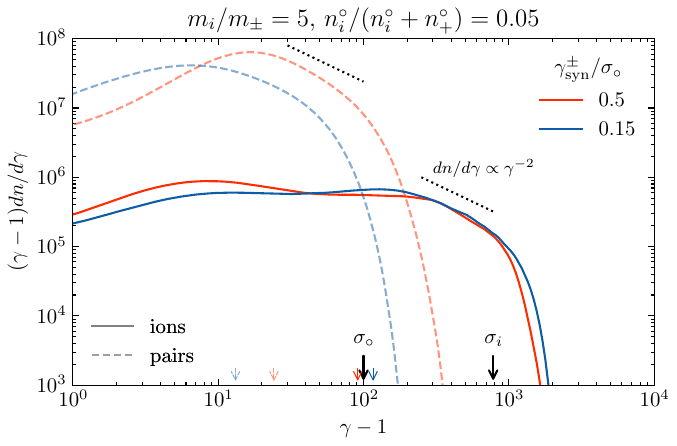}
\caption{Time-averaged energy distributions for ions (solid) and pairs (dashed) for two runs with different pair-cooling strengths. The averaging is done in a steady state, over a period of $6\lesssim ct/L\lesssim8$. The black arrows indicate $\sigma_\circ$ and $\sigma_i$ (the same for both runs), while the colored arrows indicate the average Lorentz factors of ions (solid arrows) and pairs (dashed arrows).}
\label{fig:prtl_dist}
\end{figure}

In Figure~\ref{fig:prtl_dist}, we show the energy distributions of both protons (solid) and pairs (dashed) for both of our runs. The black arrows indicate $\sigma_\circ$ and $\sigma_i$, which are the same for both of our simulations, while the colored arrows (solid and dashed) show $\langle\gamma_i\rangle$ and $\langle\gamma_\pm\rangle$ respectively. In a steady state, the pairs (dashed lines) form a spectrum that peaks at around $\gammasyn$ and drops as $\gamma^{-2}$ up to $\sim\sigma_\circ$, with the average energy being a fraction of $\gammasyn$ \citep[see][]{Chernoglazov.etal_2023}. The ions (solid lines), on the other hand, form a much harder power law\footnote{Notice that the distribution of ions is slightly steeper for the simulation with weaker synchrotron cooling for pairs (likewise, the average energy is slightly smaller). As we discussed above, this result is likely due to the limited separation of scales between the ion and pair Larmor radii. In a realistic scenario, we expect that ions will be fully decoupled from pairs, thus forming a spectrum more similar to the case with strong cooling (blue).} of around $dn_i/d\gamma \propto \gamma^{-1}$ at energies $\gamma\lesssim 200$, which transitions into a steeper distribution of $dn_i/d\gamma \propto \gamma^{-2}$ and extends roughly up to the Hillas limit, $\gammahillas\approx2000$. In a more realistic scenario applicable to the parameters of the M87* accretion flow, the power law will extend up to about $\gamma_{\rm syn}^i$, as its value is typically $\ll \gamma_H^i$. The average energy of the ions is almost insensitive to the ratio of $\gammasyn/\sigma_\circ$, and is roughly equal to $\langle\gamma_i\rangle\approx 0.15...0.2\cdot\sigma_i$ (see the dashed horizontal line in Figure~\ref{fig:meangamma_betarec_vs_t}(\emph{b})). We can also confirm our arguments, made in Section~\ref{sec:analytics}, by numerically solving the implicit equation for $\gamma_b^i$ (the position of the break), using $\gamma_H^i$ instead of $\gamma_{\rm syn}^i$. Substituting the numbers, and taking $s=2$, $\chi\approx 0.15$, we find $\gamma_b^i\approx 240$, which matches the break positions in the spectra of ions in Figure~\ref{fig:prtl_dist}.


\subsection{Radiation Spectra}
\label{sec:radiation}

In this section, we present the radiation spectra, from both pairs and ions, by postprocessing the data from our simulations. For computing the synchrotron emission spectrum, we use particle four-velocities, as well as the values of the electric and magnetic fields at the position of each particle. We use data from a snapshot at a late time step for the strongly cooled, $\gammasyn/\sigma_\circ=0.15$, simulation. 

Following our prescription from the previous sections, the synchrotron power per unit energy for a single particle with a Lorentz factor $\gamma$ (and three-velocity $\bm{\beta}$) and electric and magnetic fields at its position $\bm{E}$ and $\bm {B}$ can be written as

\begin{equation}
\label{eq:synchrotron-power}
    \frac{d\mathcal{E}_{\rm syn}^{\pm,i}}{dt d\varepsilon}\propto \tilde{b}_\perp F_{\rm syn}\left(\frac{\varepsilon}{\varepsilon_{\rm peak}^{\pm,i}}\right),
\end{equation}

\noindent where $F_{\rm syn}(\xi)\equiv \xi\int_\xi^\infty d\xi' K_{5/3}(\xi')$ is the synchrotron kernel function \citep[see, e.g.,][]{Blumenthal.Gould_1970}. Since the value of the electric field might not be negligible, we also employ the full expression for the perpendicular magnetic field component, where $\tilde{B}_\perp^2 \equiv |\bm{E}+\bm{\beta}\times\bm{B}|^2 - (\bm{\beta}\cdot \bm{E})^2$, with $|\tilde{B}_\perp| \equiv\tilde{b}_\perp B_\circ$ being the perpendicular component of the magnetic field in the frame, where $\bm{E}'\parallel \bm{B}'$ \citep[see, e.g.,][]{Cerutti.Beloborodov_2016}. We also define $\varepsilon_{\rm peak}^{\pm,i}\equiv \varepsilon_{\rm syn}^{\pm,i}\left(\gamma_{\pm,i}/\gamma_{\rm syn}^{\pm,i}\right)^2 \tilde{b}_\perp$ as the peak energy of the synchrotron emission spectrum, with $\varepsilon_{\rm syn}^{\pm,i} \equiv (3/2)\hbar (|e| B_\circ / m_{\pm,i} c) \left(\gamma_{\rm syn}^{\pm,i}\right)^2$ and $\left(\gamma_{\rm syn}^{\pm, i}\right)^2\equiv (9/4)\beta_{\rm rec}m_{\pm,i}^2 c^4/(|e|^3 B_\circ)$. Notice that the value of $\varepsilon_{\rm syn}^{\pm,i}$ does not depend on the value of $B_\circ$, and can be written as \citep[see, e.g.,][]{Hakobyan.etal_2023}:

\begin{equation}
    \varepsilon_{\rm syn}^{\pm,i}=\frac{27}{8}\frac{\beta_{\rm rec}}{\alpha_F}m_{\pm,i} c^2\approx \begin{cases}
        24~\text{MeV}\quad &\text{for }e^\pm,\\
        43~\text{GeV}\quad &\text{for ions};
    \end{cases}
\end{equation}

\noindent where $\alpha_F\equiv e^2/\hbar c\approx1/137$. Because of this, to extrapolate our simulation to realistic values, there is no need to rescale the magnetic field directly; simply rescaling the values of $\gamma_{\rm syn}^{\pm, i}$ is enough. 

Notice that in Equation \eqref{eq:synchrotron-power}, we omit the constants at the front that ultimately determine the total emitted synchrotron power. Instead, for each species, we rescale the total emitted power to be equal to the total energy deposited during reconnection into that species. This approximation is equivalent to an assumption that the observed spectrum is integrated over a period of time, longer than the characteristic cooling time of each species. For pairs, this is true because they are in the strong-cooling regime, while for ions, we implicitly assume that the integration time is comparable to the duration of the flare; in that case --- comparing equations \eqref{eq:hillas-limit-ions} and \eqref{eq:gamma-syn-ions} --- we see that the most energetic ions cool on timescales shorter than the escape time from the system (since $\gamma_H^i\gg\gamma_{\rm syn}^i$). Also, since we do not have explicit synchrotron cooling for the ions in our simulations, we instead set $\gamma_{\rm syn}^i$ close to their corresponding Hillas limit, assuming that in a realistic scenario the power-law slope of $\gamma_i^{-2}$ will extend to the burnoff limit (similar to our simulations, where it instead extends to the Hillas limit). This substitution can potentially affect the predicted emission tail beyond the burnoff limit of $43$ GeV, however, the position of the peak itself as well as the total emitted power are fully captured within the model. 

In addition to synchrotron emission, following \cite{Hakobyan.etal_2023}, we also model the IC radiation of pairs upscattering isotropically distributed low-energy (radio-to-near-IR) photons. We assume that the reconnection region is filled with a soft radiation background from the disk that has a characteristic distribution of $d\mathcal{E}_s/d\varepsilon_s\propto\varepsilon_s^\alpha$, where $\alpha\approx 0$ for $\varepsilon_s< 300$ GHz ($\approx 10^{-3}$ eV) and $\alpha\approx-1.2$ otherwise \citep{2015ApJ...809...97B}. As opposed to the synchrotron case, where we can simply rescale the $\gamma_{\rm syn}$ parameter, to properly reproduce the IC signal, relying on the unscaled soft photon background spectrum, we have to employ a realistic $e^\pm$-distribution function expected to be produced during the flaring event. For that, we use a hard-power-law slope with an exponential cutoff, $dn_\pm/d\gamma_\pm\propto\gamma_\pm^{-1}e^{-\gamma_\pm/\gamma_c^\pm}$, where we vary the value for the cutoff $\gamma_c^\pm \approx \mathcal{O}(1)\gamma_{\rm syn}^\pm$ around the synchrotron burnoff limit. The total spectrum of the emerging IC radiation can then be evaluated as

\begin{equation}
    \label{eq:ic-power}
    \frac{d\mathcal{E}_{\rm IC}^\pm}{dt d\varepsilon}\propto 
    \int \frac{dn_\pm}{\gamma_\pm^2} \int \frac{d\varepsilon_s}{\varepsilon_s^2}\varepsilon F_{\rm IC}(q,\Gamma_s),
\end{equation}

\noindent where we use the IC kernel from \cite{Blumenthal.Gould_1970}:

\begin{equation}
    F_{\rm IC}(q,\Gamma_s)\equiv 2 q \ln{q} + (1 - q)\left[(1 + 2q) + \frac{1}{2} \frac{(\Gamma_s q)^2}{1 + \Gamma q} \right],
\end{equation}

\noindent which describes individual Compton scatterings from an energy $\varepsilon_s$ to $\varepsilon$ by a particle with a Lorentz factor of $\gamma_\pm$. Here, $q\equiv (\varepsilon/\gamma_\pm m_\pm c^2)/(\Gamma_s(1-\varepsilon/\gamma m_\pm c^2))$ and $\Gamma_s\equiv 4\varepsilon_s\gamma_\pm/m_\pm c^2$. The overall normalization of the spectrum --- i.e., the total emitted power in IC --- is proportional to the energy density of soft background photons, $U_s$. The value for this parameter is not well understood, especially during the flaring event; in the quiescent state, the average value estimated from the radio-to-near-IR flux is close to $U_s\approx 10^{-2}$ erg cm${}^{-3}$  \citep{2015ApJ...809...97B}, which is roughly $0.1\%$ of $\beta_{\rm rec}U_B\approx 40$ erg cm${}^{-3}$ (for $B\approx 100$ G). For the purposes of this paper, we will employ an admittedly more optimistic value of $1\%$, which accounts for a local enhancement of soft radiation during the flare. 

\begin{figure*}[htb!]
\centering
\includegraphics[width=0.9\textwidth]{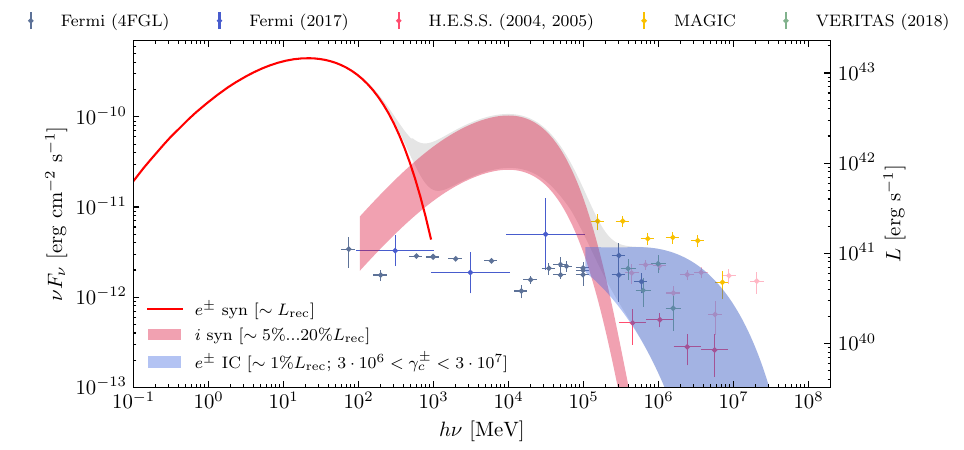}
\caption{Reconstructed multiwavelength emission of the M87* black hole during the reconnection-driven flaring event. The red line $\lesssim 100$ MeV corresponds to the synchrotron emission of strongly cooled pairs, and its luminosity is fixed at around $L_{\rm rec}$, which corresponds to the total power dissipated during the reconnection event. The synchrotron emission of protons is shown with a red band between $100$ MeV up to $\lesssim 100$ GeV, and its luminosity corresponds to our estimate from Section~\ref{sec:analytics}. Inverse-Compton emission of pairs upscattering soft photons from the disc is shown with a blue strip beyond $\gtrsim 100$ GeV; the luminosity of this component corresponds to the estimates by~\cite{Hakobyan.etal_2023}. Points with errorbars correspond to the observations performed at different times by the Fermi satellite ($\lesssim 100$ GeV), and the H.E.S.S., MAGIC and VERITAS detectors ($\gtrsim 100$ GeV)~\citep{2023arXiv230712546B, EHT_HE_M87}. Note, that our reconstructed spectrum is only valid during the strong flaring event which lasts for $\sim 10\, r_g/c$ ($\sim$ few days), while the observed GeV spectrum in this plot is integrated over a span of at least a few months, which thus includes quiescence where likely no significant flares occur.}
\label{fig:he-spectrum}
\end{figure*}

The resulting spectral components are shown in Figure~\ref{fig:he-spectrum}, where we also overplot the observed datapoints both during the quiescence as well as the flaring state: TeV data by HESS, MAGIC, and VERITAS correspond to the flaring state lasting for about a few days~\citep{EHT_HE_M87}, while the GeV data from Fermi (4FGL) are integrated over a span of three months in 2017~\citep{EHT_HE_M87} and a few years \citep{2023arXiv230712546B}. With the labels, we highlight the ranges of values for different parameters used to produce a range of predictions for the ion-synchrotron and pair-IC signals (with respect to the luminosity of the synchrotron emission of pairs, which is fixed at around $L_{\rm rec}$). As expected, the synchrotron spectra for both pairs and protons peak at around $\varepsilon_{\rm syn}^{\pm,i}$. The IC signal reaches energies of the order of $\mathcal{O}({10})\left(\gamma_c^\pm\right)^2\varepsilon_s\approx 1$ TeV. 

Finally, it is important to emphasize that the relative normalizations of different spectral components presented in this work should not be taken at face value and should be thought as order-of-magnitude estimates based on our best understanding of the microphysics. There are two main reasons why the actual observed luminosities may differ from those predicted here. First, our discussion focuses on the localized microphysical picture of the flare, while in reality bulk motions within the reconnection region may strongly affect the observed luminosity of the signal due to Doppler boosting. On top of that, different components are most likely very anisotropic and are thus beamed in different directions; for instance, as was found by \cite{Chernoglazov.etal_2023} and \cite{solanki2025}, the synchrotron emission of strongly cooled pairs is likely oriented along the upstream magnetic field (the $x$-direction in our simulations), whereas the synchrotron emission of weakly cooled protons likely coincides with their dominant direction of motion perpendicular to the sheet (the $y$-direction).

\section{Conclusions} 
\label{sec:conclusions}

Episodic magnetic reconnection during MAD accretion states can give rise to efficient $\gamma$-ray emission by particles accelerated in the reconnection zone. At sufficiently low densities, the magnetization upstream of the reconnecting current sheet is regulated by electron-positron pair creation, owing to the annihilation of MeV photons generated through rapid synchrotron cooling of the accelerated pairs.  This MeV emission taps the majority of the dissipated power.  A small fraction of the dissipation power is released as TeV photons through IC scattering of ambient soft radiation off the energetic  pairs. It has been shown previously \citep{Hakobyan.etal_2023} that the rapid TeV flares detected in M87 can be produced by this process during flux eruption states.   However, the spectra computed in \cite{Hakobyan.etal_2023} indicate that pair radiation alone cannot account for the variable GeV emission detected in M87 by the Fermi observatory. 

In this paper, we argue that GeV emission can be naturally produced by ion-synchrotron radiation.  We have shown analytically that under the conditions expected in the inner magnetosphere during flux eruption states, ions accelerated in the current sheet will emit at the synchrotron burnoff limit, $\sim 40$ GeV, before escaping the system, implying high radiative efficiency.  Using 3D radiative PIC simulations of ion-pair plasma we then computed the ion energy distribution and the energy partition between the accelerated ions and pairs, finding that the GeV luminosity radiated by the ions constitutes a few percent of the reconnection power. We have also demonstrated that IC emission by the accelerated pairs can explain most of the TeV flaring activity \citep[see][]{Hakobyan.etal_2023}, although this may be challenged by the most extreme flares \citep[e.g., see the 2017 flare reported by MAGIC;][]{EHT_HE_M87}.  
We stress that our analysis  focuses on the localized microphysical picture of the flare and neglects global effects, such as bulk motions within the reconnection zone, photon lensing, and anisotropies, which might alter the relative normalization of the different spectral components \citep{solanki2025}.  

Our 3D PIC simulations of pair-ion reconnection also deliver a few general results, which hold regardless of the specific astrophysical system. First, we have shown that the synchrotron-cooled pairs form a narrow layer containing most of the particles, while the high-energy ions occupy a broader diffuse layer, where ion pressure gradients balance electromagnetic stresses. Second, while it is essential for the overall force balance, ion feedback does not seem to affect the value of the reconnection rate. Third, for the conditions explored in this paper, the mean ion Lorentz factor saturates at $\gamma_b\sim \sigma_i/6$ ($\sigma_i$ is the ion magnetization), regardless of the strength of the pair cooling, as long as the dynamics of the protons is sufficiently decoupled from that of the pairs. Finally, the ion energy spectrum can be modeled as a broken power law, with a hard slope $f_i\propto \gamma^{-1}$ below the break, $\gamma_b$, and a softer tail of $f_i\propto \gamma^{-2}$ beyond it. The break energy is of the order of the mean energy per ion, which then scales $\propto \sigma_i$. A broken-power-law proton spectrum with break energy $\propto \sigma_i$ has also been employed by reconnection-based models of the TeV neutrinos from NGC 1068 \citep{fiorillo_24,karavola_25}.
While our results are based on simulations having zero guide field, future work will need to assess how the properties of the proton spectrum in pair-proton reconnection depend on the guide-field strength. 

\section*{acknowledgment}

H.H. thanks Alexander Chernoglazov for insightful discussions and valuable feedback. \fixed{The authors would also like to thank the anonymous referee for the insightful comments that helped to improve the quality of this work.} This work was supported by a grant from the Simons Foundation (MP-SCMPS-00001470). 
A.L. and A.P. acknowledge support from a BSF grant (2022314). A.L. wishes to thank Frank Rieger for enlightening discussions and help, and the Canadian Institute for Theoretical Astrophysics for their warm hospitality and support. L.S. acknowledges support from DoE Early Career Award DE-SC0023015, NASA ATP 80NSSC24K1238, NASA ATP 80NSSC24K1826, and NSF AST-2307202. A.P. additionally acknowledges support from NASA grant 80NSSC22K1054, an Alfred P.~Sloan Research Fellowship, and a Packard Foundation Fellowship in Science and Engineering. This research was facilitated by the Multimessenger Plasma Physics Center (MPPC), NSF grant PHY-2206607.
B.R. acknowledges support from the Natural Sciences \& Engineering Research Council
of Canada (NSERC) and the Canadian Space Agency (23JWGO2A01). B.R. acknowledges a guest researcher
position at the Flatiron Institute, supported by the Simons Foundation. 


\bibliography{references,hh-refs}{}
\bibliographystyle{aasjournalv7}

\end{document}